\documentclass[twocolumn,10pt,superscriptaddress,times]{revtex4-2}
\usepackage{graphics,graphicx,epsfig,ulem,epstopdf,bm,longtable,url,geometry,datetime,physics}
\usepackage[colorlinks=true,linkcolor=blue,urlcolor=blue,citecolor=blue]{hyperref}
\usepackage{amsmath,amssymb,latexsym,float,amsfonts,hyperref,color,setspace,overpic}
\usepackage{microtype,comment} 
\usepackage[caption=false]{subfig} 
\usepackage[ansinew]{inputenc}
\usepackage[usenames,dvipsnames]{pstricks}
\usepackage[pagewise,mathlines]{lineno}

\def\footnoterule{\kern -1mm \hrule width 6.0cm \kern 2.2mm}%
\usepackage{tikz,xcolor,hyperref}
\definecolor{lime}{HTML}{A6CE39}
\DeclareRobustCommand{\orcidicon}{%
    \begin{tikzpicture}
    \draw[lime, fill=lime] (0,0)
    circle [radius=0.16]
    node[white] {{\fontfamily{qag}\selectfont \tiny ID}};\draw[white, fill=white] (-0.0625,0.095)
    circle [radius=0.007];
    \end{tikzpicture}
    \hspace{-2mm}}
\foreach \x in {A, ..., Z}
{\expandafter\xdef\csname orcid\x\endcsname{\noexpand\href{https://orcid.org/\csname orcidauthor\x\endcsname}{\noexpand\orcidicon}}}

\begin{document}
\title{Lamb-shift-mediated energy transfer in open quantum batteries}
 
\author{Liang Luo}
\affiliation{Center for Quantum Materials and Computational Condensed Matter Physics, Faculty of Science, Kunming University of Science and Technology, Kunming, 650500, PR China}
\author{Shun-Cai Zhao\orcidA{}}
\email[Corresponding author: ]{zsczhao@126.com.}
\affiliation{Center for Quantum Materials and Computational Condensed Matter Physics, Faculty of Science, Kunming University of Science and Technology, Kunming, 650500, PR China}
\begin{abstract}
Open quantum batteries (QBs) operate under unavoidable system--environment interactions, where both dissipation and coherent frequency renormalization can
affect their dynamics. While dissipative effects have been extensively studied, the role of environment-induced frequency shifts, such as the Lamb shift, remains
less explored. Here, we investigate a driven open QB consisting of two coherently coupled quantum harmonic oscillators representing the charger and
the battery. By incorporating dissipation and Lamb-shift corrections within a Lindblad master equation, we show that the Lamb shift renormalizes the system
eigenfrequencies and modifies the resonance condition with the external drive. We further demonstrate that the resulting frequency renormalization leads to a
mode-selective energy-transfer process, producing a redistribution of energy between the charger and the battery. This behavior is characterized through a
supermode decomposition of the coupled system, which reveals how the environment-induced frequency shift alters the dominant energy-transfer channel.
Our results clarify the role of coherent environmental effects in open quantum batteries and provide a physical framework for understanding work-extraction
dynamics beyond purely dissipative descriptions.
\end{abstract}
\date{\today}
\maketitle

\section{Introduction}\label{sec:Introduction}

Quantum batteries (QBs) represent a cutting-edge frontier in energy storage technology \cite{PhysRevLett.134.180401,PhysRevA.100.043833,d9k1-75d4,PhysRevE.111.044118,PhysRevA.111.042216,p93y-jflt,xqtv-qbyk}, promising to revolutionize next-generation quantum devices\cite{Quach2022,56-PENG2026131462} by leveraging non-classical resources such as coherence\cite{PhysRevE.102.052109}, entanglement\cite{PhysRevA.104.L030402}, and collective quantum effects \cite{PhysRevA.105.062203}. Unlike traditional electrochemical cells, QBs aim to achieve ultrafast charging rates and high power densities by exploiting the unique thermodynamic laws at the microscale \cite{PhysRevE.102.062133,PhysRevLett.134.180401,doi:10.1126/sciadv.adw8462}. The core research motivation lies in developing efficient and stable quantum energy harvesters capable of powering nanoscale quantum computing and communication networks\cite{PhysRevLett.118.150601,PhysRevLett.120.117702}.

The evolution of QB research can be broadly categorized into two stages. Early investigations primarily focused on closed quantum systems\cite{PhysRevLett.129.130602,PhysRevA.97.022106,PhysRevE.99.052106}, where the battery's evolution is assumed to be unitary and isolated from any environment. These studies established the foundational bounds for work extraction and charging power\cite{PhysRevE.87.042123,PhysRevE.100.032107,PhysRevB.99.205437}. More recently, attention has shifted towards the study of open quantum batteries\cite{PhysRevE.104.044116,PhysRevLett.134.130401,PhysRevE.104.054117,PhysRevA.102.052223}, which accounts for the realistic interactions between the system and its surrounding reservoir. Within this context, significant effort has been devoted to mitigating dissipative losses and decoherence effects\cite{PhysRevApplied.14.024092,ABAKUMOVA2023169322,PhysRevLett.122.210601,PhysRevA.107.022214}. In continuity with our previous work on the dynamics of open quantum systems \cite{Zhong2021,Zhu2023}, we have observed that environmental coupling does more than just induce decay; it fundamentally reshapes the system's energy landscape.
 
However, a critical gap persists in the existing literature: the vast majority of prior open-system QB studies\cite{PhysRevB.99.035421,PhysRevE.102.062133,PENG2026131462,Carrega2020Dissipative} have failed to consider system-environment interference, effectively neglecting the Lamb shift\cite{PhysRev.72.241,TheoryofOpen1}. By assuming that the environment only induces energy dissipation (represented by decay rates) without modulating the bare system frequencies, these models oversimplify the resonance conditions. Such an oversight is particularly problematic in experimental platforms like circuit QED or cavity QED, where vacuum fluctuations inevitably induce frequency shifts \cite{Fragner2008}. Although dissipative effects have been extensively investigated in open quantum batteries, the coherent frequency renormalization induced by the environment, such as the Lamb shift, has received comparatively less attention. Such renormalization can become relevant in precisely controlled platforms, where small shifts of system frequencies may modify resonance conditions.

In this work, we aim to bridge this gap by explicitly incorporating Lamb-shift corrections\cite{PhysRevApplied.23.024010,Fragner2008,PhysRevB.99.035421} into the charging dynamics of an open QB system. We model both the charger and the battery as quantum harmonic oscillators\cite{PhysRevA.93.052119,PhysRevLett.123.080602} and derive a Lindblad master equation that accounts for both dissipation and the interference-induced frequency renormalization. By explicitly incorporating Lamb-shift corrections into the dynamics, this work examines how environment-induced frequency renormalization modifies the resonance structure and energy-transfer pathways of an open quantum battery\cite{PhysRev.72.339, PhysRevA.81.052501}.

Our results reveal a mode-selective energy-transfer behavior arising from the Lamb-shift-modified resonance condition. By tuning the driving frequency relative to the renormalized eigenfrequencies, the energy distribution between the charger and the battery can be modified. This behavior is referred to as a switching-like redistribution of energy between different storage channels. Furthermore, we propose a frequency correction strategy based on the Lamb-shifted eigenvalues to restore ergotropy that would otherwise degrade due to environmental interference. These findings suggest a possible mechanism for controlling energy redistribution in open quantum batteries.

The remainder of this paper is organized as follows: In Sec.\ref{sec:model}, we describe the model setup and derive the master equation with Lamb-shift corrections. In Sec.\ref{sec:results}, we present the numerical results and the physical analysis of the switchable charging dynamics. Finally, the experimental feasibility and conclusions are summarized in Sec. \ref{sec:Experimental Feasibility} and Sec. \ref{sec:Conclusion}.

\section{Model and Theoretical Framework}\label{sec:model}

\begin{figure}[htb]
  \centering
  \includegraphics[width=0.3\textwidth]{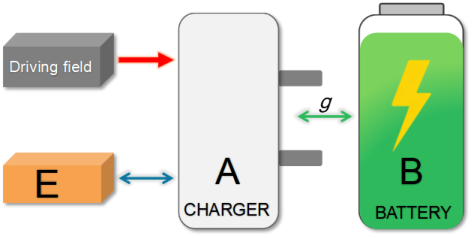}
  \caption{Schematic of the quantum battery(QB) system. The charger (middle) and battery (right) are quantum harmonic oscillators coupled with strength $g$ (green arrow). The external driving field (left) initiates the charging process (red arrow), while interaction with the environment $E$ leads to both dissipation and an interference-induced shift (blue  bidirectionalarrow) between the charger and the thermal reservoir.}
  \label{model}%
\end{figure}

As illustrated schematically in Fig.~\ref{model}, we consider an open quantum battery (QB) composed of a \textit{charger} ($A$) and a \textit{battery} ($B$), both modeled as quantum harmonic oscillators (QHOs) with eigenfreencies $\omega_a$ and $\omega_b$, respectively~\cite{PhysRevB.98.205423}. The charger is coherently driven by an external laser field of amplitude $F$ and frequency $\omega_f$, while transferring energy to the battery through a bilinear interaction with coupling strength $g$. Throughout this work, we set $\hbar=1$. Under the dipole and rotating-wave approximations (RWA)~\cite{LoudonLight,53-xqtv-qbyk}, the total Hamiltonian is
\begin{eqnarray}\label{1}
\hat{H} =& \omega_a \hat{a}^\dagger \hat{a} + \omega_b \hat{b}^\dagger \hat{b} + g \left(\hat{a}\hat{b}^\dagger + \hat{b}\hat{a}^\dagger \right) \nonumber\\
         &+ F \left( e^{i\omega_f t} \hat{a} + e^{-i\omega_f t} \hat{a}^\dagger \right),
\end{eqnarray}
where $\hat{a}^\dagger$ ($\hat{a}$) and $\hat{b}^\dagger$ ($\hat{b}$) denote the creation (annihilation) operators of the charger and battery, respectively.

We assume that the battery remains well isolated, whereas the charger is coupled to a thermal reservoir inducing both dissipation and environment-induced frequency renormalization. The dynamics of the total density matrix $\hat{\rho}_{AB}$ obey the Lindblad master equation~\cite{TheoryofOpen1}
\begin{eqnarray}\label{2}
\dot{\hat{\rho}}_{AB} = &-i \left[\hat{ H}, \hat{\rho}_{AB} \right] + \gamma_a \left( N(T) + 1 \right) \mathcal{D}_{\hat{a}} \left[ \hat{\rho}_{AB} \right] \nonumber\\
& + \gamma_a N(T) \mathcal{D}_{\hat{a}^\dagger} \left[ \hat{\rho}_{AB} \right]
- i\Delta_{L} \left[ \hat{a}^\dagger \hat{a}, \hat{\rho}_{AB} \right],
\end{eqnarray}
where $\gamma_a$ is the decay rate and $\mathcal{D}_{\hat{c}}[\hat{\rho}] = \hat{c}\hat{\rho} \hat{c}^\dagger - \frac{1}{2}\{ \hat{c}^\dagger \hat{c}, \hat{\rho} \}$ is the standard dissipator. The mean thermal occupation is $N(T)=[\exp(\omega/k_B T)-1]^{-1}$. The Lamb shift $\Delta_L$ originates from the dispersive component of the system--environment interaction and appears as a coherent renormalization of the system Hamiltonian. Unlike dissipative terms that directly modify populations and coherences, the Lamb shift changes the effective transition frequencies of the coupled system. Therefore, in the present hybrid quantum battery, it modifies the resonance condition between the driving field and the collective modes, thereby influencing the transient energy-transfer dynamics.

To establish the microscopic origin of both $\gamma_a$ and $\Delta_L$, we model the charger as linearly coupled to a structured bosonic reservoir characterized by the Ohmic spectral density with Lorentz-Drude cutoff $\omega_c$\cite{PhysRevA.108.012210,PhysRevA.104.052203,PhysRevA.96.062108,PhysRevA.77.032117,PhysRevA.101.042130},
The choice of a local master equation is motivated by the fact that the environmental coupling is restricted to the charger subsystem, while the battery remains isolated from the reservoir. For coupled subsystems with comparable coherent coupling and dissipation strengths, a fully secular global treatment may suppress non-secular
contributions associated with the hybridized modes. Such contributions can be relevant for describing transient energy-transfer dynamics.

In the present model, the Hamiltonian is quadratic and the external driving is linear in the bosonic operators. Within the Markovian regime, the first-order moments obtained from the local master equation\cite{Blondel2014,Hofer2017,PhysRevA.102.052208} reproduce the corresponding linear quantum Langevin dynamics. Therefore, the Lamb-shift correction considered here consistently describes the environment-induced frequency renormalization within the adopted open-system framework. To characterize the environmental influence, we consider an Ohmic reservoir\cite{Carrega2020Dissipative,RevModPhys.96.031001} with a Lorentz-Drude cutoff spectral density, which determines both the dissipative rates and the Lamb-shift contribution. The corresponding spectral density is given by

\begin{equation}\label{3}
J(\omega)=\frac{\eta}{\pi}\omega
\frac{\omega_c^2}{\omega^2+\omega_c^2},
\end{equation}
where $\eta$ denotes the dimensionless system-bath coupling strength and $\omega_c$ is the cutoff frequency of the reservoir.

Within the Born-Markov approximation, the real and imaginary parts of the reservoir correlation function determine dissipation and coherent renormalization, respectively. Consequently, $\gamma_a$ and $\Delta_L$ are microscopically linked through the Kramers-Kronig relations\cite{PhysRevApplied.22.L041002,PhysRev.161.143,PhysRevD.108.044015,PhysRevB.103.144303},
\begin{subequations}\label{4}
\begin{align}
\gamma_a &= 2\pi J(\omega_a)=2\eta\omega_a\frac{\omega_c^2}{\omega_a^2+\omega_c^2}, \label{4a}\\
\Delta_L &= \mathcal{P}\int_0^\infty \frac{J(\omega')}{\omega_a-\omega'}d\omega',  \label{4b}
\end{align}
\end{subequations}
where $\mathcal{P}$ denotes the Cauchy principal value. The magnitude and sign of $\Delta_L$ depend on the spectral distribution of the reservoir and the system--bath detuning. Therefore, varying $\Delta_L$ in the following analysis can be interpreted as exploring different structured environmental conditions rather than introducing an independent external control parameter. Evaluating the integral analytically (see Appendix \ref{appendix}) yields
\begin{equation}\label{5}
\Delta_L=\frac{\eta\omega_c^2}{\pi (\omega_a^2+\omega_c^2)}
\left[\omega_a\ln\left(\frac{\omega_a}{\omega_c} \right)-\frac{\pi}{2}\omega_c\right].
\end{equation}

Importantly, the ratio $\Delta_{L}/\gamma_{a}$ derived from this single Lorentzian-broadened Ohmic reservoir is expressed as
\begin{equation}\label{6}
\frac{\Delta_L}{\gamma_a}
=\frac{1}{2\pi}
\left[
\ln\left( \frac{\omega_a}{\omega_c}\right)
-\frac{\pi\omega_c}{2\omega_a}
\right].
\end{equation}

To avoid ambiguity regarding frequency renormalization, we explicitly define $\omega_a$ as the bare frequency of the charger subsystem prior to coupling with the dynamical bath variation. In the spirit of the Caldeira-Leggett framework, the divergent cutoff-dependent term arising from the system-environment interaction is absorbed via a standard potential counter-term, rendering the residual Lamb shift $\Delta_L$ in Eq.~\eqref{4b} a finite, physically observable frequency shift. Experimentally, the baseline physical frequency measured in a static environment is $\omega_{a,\mathrm{phys}} $=$ \omega_a + \Delta_L^{(0)}$. Variations in the structured environment (e.g., modulating the coupling strength $\eta$ or spectral cutoff $\omega_c$) induce a controllable frequency shift relative to this baseline, $\delta \Delta_L $=$ \Delta_L - \Delta_L^{(0)}$. Consequently, identifying $\Delta_L$ in our dynamical equations represents a physical modulation without double-counting the environment-induced renormalization.

For structured reservoirs with $\omega_c \sim 5\omega_a$--$10\omega_a$, the ratio $\Delta_L/\gamma_a$ naturally falls within $0.5$--$5.0$, confirming that the parameter regime explored here is both physically accessible and microscopically self-consistent. This feature also distinguishes the vacuum-induced self-gating mechanism from trivial external frequency tuning.

It is worth noting that at zero temperature ($N(T)=0$), the Heisenberg equations of motion for the expectation values are linear. Under exact resonance $\omega_f = \lambda_{\pm}$, the steady-state amplitude of the charger simplifies to $|A|_{\text{ss}} = 2F/\gamma_a$, which is asymptotically independent of $\Delta_L$ and $g$. However, quantum batteries operate inherently as non-equilibrium transient energy storage devices, where the figure of merit is the peak extractable work (ergotropy) \cite{Strasberg2016,Francica2017,PhysRevLett.122.047702} achieved prior to full dissipation. The environment-induced Lamb shift $\Delta_L$ dynamically dictates the transient energy transfer pathways and peak charging power, even though the asymptotic steady state remains governed by the drive-to-dissipation ratio.
Equation~\eqref{6} reveals that for a standard broadband thermal or vacuum reservoir with $\omega_{c} \gtrsim \omega_{a}$, the single-reservoir microscopic calculation inherently yields a negative Lamb shift ($\Delta_{L} < 0$, i.e., a redshift). For the Ohmic-type reservoir considered above, the microscopic calculation mainly yields a negative Lamb shift. Positive shifts can be realized in structured reservoirs with modified spectral distributions. Therefore, both signs of $\Delta_L$ are considered below to characterize the general influence of environment-induced frequency renormalization. In the following dynamical analysis, we explore both negative and positive values of $\Delta_{L}$ to establish a comprehensive physical picture of the environment-gated energy redistribution.

The expectation value of an arbitrary operator $\hat{A}$ evolves according to the adjoint master equation~\cite{Carmichael1993Lecture1},
\begin{eqnarray} \label{7}
\frac{d}{dt}\langle\hat{A}\rangle =& -i\langle[\hat{A},\hat{H}]\rangle + \gamma_{a}(N(T)+1)\mathcal{D}[\hat{a}]\langle\hat{A}\rangle \nonumber \\
                                   & + \gamma_{a}N(T) \mathcal{D}[\hat{a}^\dagger]\langle\hat{A}\rangle
                                   - i\Delta_{L}\langle[\hat{A},\hat{a}^{\dagger}\hat{a}]\rangle,
\end{eqnarray}
where
$\mathcal{D}[\hat{O}]\langle\hat{A}\rangle=\langle \hat{O}^{\dagger}\hat{A}\hat{O}-\frac{1}{2}\{\hat{O}^{\dagger}\hat{O},\hat{A}\}\rangle$.
In the zero-temperature limit [$N(T)=0$], the dynamics reduce to~\cite{PhysRevB.99.035421}

\begin{subequations}\label{8}
\begin{align}
\langle \dot{\hat{a}} \rangle &= -i F e^{-i\omega_f t}-i g \langle \hat{b} \rangle-\frac{\Gamma}{2} \langle \hat{a} \rangle,\label{8a}\\
\langle \dot{\hat{b}} \rangle &=-i g \langle \hat{a} \rangle-i \omega_b \langle \hat{b} \rangle,   \label{8b}
\end{align}
\end{subequations}
with the complex decoherence parameter $\Gamma/2 $=$ \gamma_a/2+i(\omega_a+\Delta_L)$.
 
Before introducing the extractable work, we clarify the thermodynamic meaning of the quantity used throughout this work. 
For a general continuous-variable Gaussian state\cite{PhysRevLett.133.260401,PhysRevA.96.042341,LOPEZSALDIVAR2023128676,PhysRevA.111.052404,43jm-qkz2}, the total ergotropy depends not only on the first-order moment $\langle \hat{o}_i\rangle$, but also on the covariance matrix associated with thermal and squeezing fluctuations\cite{PhysRevA.47.4487,43jm-qkz2}. Accordingly, the total ergotropy can generally be decomposed as
\begin{equation}\label{aa}
W_{\mathrm{total}}=W_D+W_{\mathrm{cov}},
\end{equation}
where $W_D$ denotes the coherent displacement contribution and $W_{\mathrm{cov}}$ arises from non-passive covariance structures.

In the present model, the Hamiltonian is bilinear and the external driving is linear in the bosonic operators. 
For the zero-temperature limit $N(T)=0$, which is adopted in the main simulations, an initial coherent Gaussian state remains a displaced vacuum Gaussian state throughout the Lindblad evolution. Consequently, the covariance matrix retains the passive vacuum form and does not contribute to extractable work, yielding
\begin{equation}\label{bb}
W_{\mathrm{total}}=W_D
=\omega_i|\langle \hat{o}_i(t)\rangle|^2.
\end{equation}

For finite temperatures $N(T)\neq0$, thermal fluctuations introduce additional passive mixedness, and the above expression no longer represents the total ergotropy. In this regime, Eq.~\eqref{bb} should therefore be interpreted specifically as the coherent displacement ergotropy associated with first-order coherent excitation.
 
To characterize the charging performance, we evaluate the coherent displacement ergotropy, namely the maximum extractable work associated with coherent displacement under cyclic unitary operations~\cite{Allahverdyan2004}. For QHO subsystems initially prepared in the ground state, the instantaneous coherent displacement ergotropy of the charger ($W_{D,A}$) and battery ($W_{D,B}$) is

\begin{align}\label{9}
W_{D,i}(t) =E_i(t)-\min_{\hat{U}}\mathrm{Tr}\left[\hat{H}_i\hat{U}\hat{\rho}_i(t)\hat{U}^\dagger\right]
=\omega_i |\langle \hat{o}_i(t)\rangle|^2,
\end{align}
where $i\in\{A,B\}$ and $\hat{o}\in\{\hat{a},\hat{b}\}$. Here, $\hat{H}_i=\omega_i\hat{o}^\dagger\hat{o}$ is the free Hamiltonian of the corresponding subsystem and $\hat{\rho}_i(t)$ is the reduced density matrix. By numerically solving Eqs.~\eqref{8} and \eqref{9}, we characterize the charging and energy-storage dynamics of the QB in the presence of environmental effects.
\section{Results and Discussions}\label{sec:results}
\subsection{Impact of Lamb shift on energy distribution}

The interaction between the charger and the reservoir induces a Lamb-shift term
$-i\Delta_L[\hat{a}^\dagger\hat{a},\hat{\rho}_{AB}]$ in Eq.~\eqref{2}, arising from vacuum fluctuations of the structured environment. This contribution renormalizes the coherent evolution of the charger mode and modifies the detuning between the external drive and the system eigenmodes $\lambda_{\pm}$, thereby affecting the energy distribution dynamics.

To isolate this effect from externally imposed detuning, we consider two driving protocols for the frequency $\omega_f$:

\begin{itemize}
    \item[\text{(i)}] 
   \textit{Fixed-drive protocol(Protocol I):}  The driving frequency $\omega_f$ is selected prior to evolution (e.g., set to the bare or unperturbed resonance) and kept strictly time-invariant. As the environmental spectral parameters change, the Lamb shift $\Delta_L$ shifts the underlying eigenfrequencies $\lambda_{\pm}$, thereby dynamically modifying the effective drive-system detuning without any manual readjustment of the laser.

    \item[\text{(ii)}] 
  \textit{On-resonance protocol (Protocol II):} For a given fixed environmental configuration with a known $\Delta_L$, the driving frequency is pre-set to match the renormalized eigenmode $\omega_f = \lambda_{\pm}$ to evaluate the theoretical upper bound of the resonant energy transfer. 
\end{itemize}

To clarify the operational nature of these protocols and address the interplay between parameters, several physical remarks are in order. In Protocol I (Fixed-drive protocol), the driving frequency $\omega_f$ is held strictly constant throughout the dynamic evolution. As environmental variations alter $\Delta_L$, the system's internal energy spectrum renormalizes, which dynamically alters the effective drive-mode detuning without any real-time active feedback or active readjustment of the external laser. This process represents an autonomous, environment-gated switching mechanism. Conversely, Protocol II serves strictly as a benchmark protocol to probe the maximum achievable resonant energy transfer capacity under a modified environmental configuration. Furthermore, while $\gamma_a$ and $\Delta_L$ are microscopically coupled via the Kramers-Kronig relation [Eqs.~\eqref{4a}-\eqref{4b}], sweeping $\Delta_L$ at a constant baseline $\gamma_a$ in our numerical simulations serves as a controlled methodology to isolate the coherent effect of frequency renormalization from pure dissipative dampening. Such a regime is physically realized in structured reservoirs where local spectral slope engineering modulates the principal-value integral $\Delta_L$ while preserving the local spectral density $J(\omega_a)$ at the subsystem frequency.

These two protocols provide a controlled way to distinguish Lamb-shift-induced renormalization from externally tuned detuning effects.

In the system Hamiltonian, the Lamb shift can be incorporated by comparing the renormalized coherent term with the free Hamiltonian contribution
$-i[\omega_a\hat{a}^\dagger\hat{a},\hat{\rho}_{AB}]$~\cite{Carmichael1993Lecture1},
which leads to the effective frequency shift
\begin{equation}\label{10}
\omega_a'=\omega_a+\Delta_L.
\end{equation}

Using this renormalized frequency, the effective Hamiltonian takes the form
\begin{equation}\label{11}
\begin{aligned}
\hat{H}^{\prime}=&\omega_a^{\prime}\hat{a}^\dagger \hat{a}+\omega_b\hat{b}^\dagger \hat{b}
+g\left(\hat{a}\hat{b}^\dagger+\hat{b}\hat{a}^\dagger\right)\\
&+F\left(e^{i\omega_ft}\hat{a}+e^{-i\omega_ft}\hat{a}^\dagger\right).
\end{aligned}
\end{equation}

This renormalized form explicitly incorporates the environment-induced modification of the energy spectrum. For two coupled bosonic modes, Eq.~\eqref{11} can be rewritten as~\cite{WU2022127779}

\begin{equation}\label{12}
\hat{H}^{\prime}=\begin{pmatrix}
\hat{a}^\dagger &
\hat{b}^\dagger
\end{pmatrix}
\mathbf{G}
\begin{pmatrix}
\hat{a}\\
\hat{b}
\end{pmatrix}
+F\left(e^{i\omega_ft}\hat{a}+e^{-i\omega_ft}\hat{a}^\dagger\right),
\end{equation}
where
\(\mathbf{G}=\begin{pmatrix}
\omega_a' & g\\
g & \omega_b
\end{pmatrix}
\).
The corresponding eigenfrequencies are

\begin{equation}\label{13}
\lambda_{\pm}=\frac{\omega_a'+\omega_b}{2}\pm\sqrt{\left(\frac{\omega_a'-\omega_b}{2}\right)^2+g^2}.
\end{equation}

In the absence of environmental renormalization ($\Delta_L=0$) and under resonance ($\omega_a=\omega_b=\omega$), Eq.~\eqref{13} reduces to the standard normal-mode frequencies
$\lambda_{\pm}=\omega\pm g$. Including the Lamb shift modifies the resonance peaks to

\begin{equation}\label{14}
\lambda_{\pm}=\omega+\frac{\Delta_L}{2}\pm\sqrt{g^2+\left(\frac{\Delta_L}{2}\right)^2}.
\end{equation}

Hence, the optimal driving condition is no longer determined by the bare frequencies, but by the environment-renormalized spectrum.

\begin{figure*}
\centering
\subfloat
{
\begin{overpic}[width=0.22\textwidth]{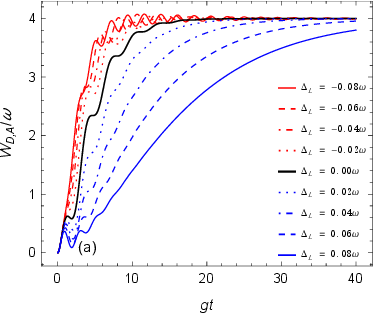}
\end{overpic}
\captionsetup{labelformat=empty}
\label{fig2a}
}
\subfloat
{
\begin{overpic}[width=0.22\textwidth]{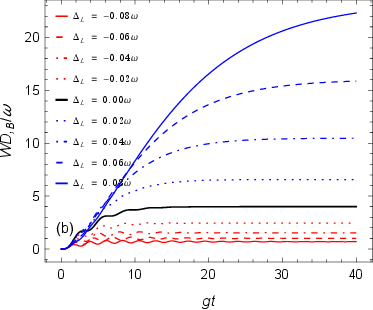}
\end{overpic}
\captionsetup{labelformat=empty}
\label{fig2b}
}
\subfloat
{
\begin{overpic}[width=0.22\textwidth]{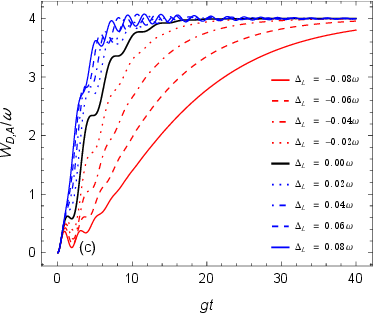}
\end{overpic}
\captionsetup{labelformat=empty}
\label{fig2c}
}
\subfloat
{
\begin{overpic}[width=0.22\textwidth]{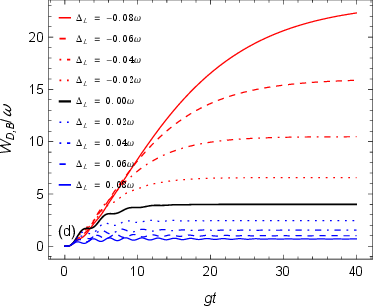}
\end{overpic}
\captionsetup{labelformat=empty}
\label{fig2d}
}
\caption{Evolution of the charger energy $W_{D,A}$ and battery coherent displacement ergotropy $W_{D,B}$ under weak coupling ($g=0.04\omega$).
For (a),(b), the driving frequency is fixed at the unperturbed eigenfrequency $\lambda_-(\Delta_L=0)$, while for (c),(d) it is fixed at $\lambda_+(\Delta_L=0)$.
Other parameters are $F=0.05\omega$, $\gamma_a=0.05\omega$, and $N(T)=0$.
}\label{fig1}
\end{figure*}

Fig.~\ref{fig1} shows the evolution of the charger energy $W_{D,A}$ and coherent displacement ergotropy $W_{D,B}$ under weak coupling. A pronounced switching behavior emerges around the critical point $\Delta_L=0$, separating two distinct dynamical regimes.

This switching originates from the Lamb-shift-induced displacement of the eigenfrequencies, which alters the resonance condition between the drive and the hybridized modes. As $\Delta_L$ changes sign, the alignment between $\omega_f$ and $\lambda_{\pm}$ is reversed, thereby redistributing the dominant energy-transfer channel [see Figs.~\ref{fig1}(a)--\ref{fig1}(d)].

It should be noted that while the single Ohmic-type continuum reservoir explicitly models the $\Delta_{L} < 0$ regime (redshift), the complementary $\Delta_{L} > 0$ regime (blueshift) naturally arises when the system mode is coupled to structured reservoirs exhibiting band-gap edges or inverted detuning profiles. The symmetric switching behavior observed across $\Delta_{L} = 0$ in Figs.~2--4 demonstrates how the sign and magnitude of the environment-induced shift dictate the alignment between the driving field and the hybridized supermodes.

For $\omega_f=\lambda_{-}$, a redshift ($\Delta_L<0$) enhances energy release from the charger [Fig.~\ref{fig1}(a)], whereas a blueshift ($\Delta_L>0$) favors ergotropy accumulation in the battery [Fig.~\ref{fig1}(b)]. The trend reverses for $\omega_f=\lambda_{+}$ [Figs.~\ref{fig1}(c) and \ref{fig1}(d)], reflecting an exchange of dominant weights between the two normal modes.

These results identify the Lamb shift as an effective environment-controlled switching knob. By jointly selecting the driving mode and the induced frequency renormalization, one may selectively optimize either energy release or work extraction.

\begin{figure*}
\centering
\subfloat
{
\begin{overpic}[width=0.22\textwidth]{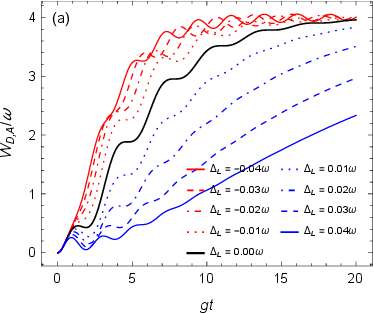}
\end{overpic}
\captionsetup{labelformat=empty}
\label{fig1a}
}
\subfloat
{
\begin{overpic}[width=0.22\textwidth]{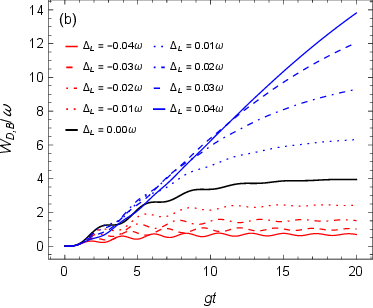}
\end{overpic}
\captionsetup{labelformat=empty}
\label{fig1b}
}
\subfloat
{
\begin{overpic}[width=0.22\textwidth]{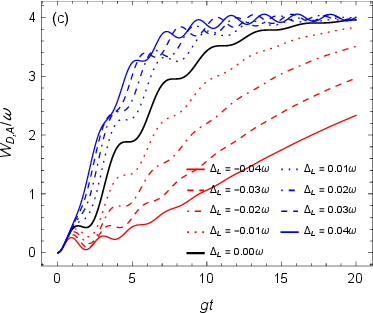}
\end{overpic}
\captionsetup{labelformat=empty}
\label{fig1c}
}
\subfloat
{
\begin{overpic}[width=0.22\textwidth]{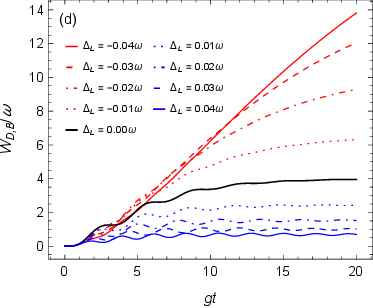}
\end{overpic}
\captionsetup{labelformat=empty}
\label{fig1d}
}
\caption{Evolution of the charger energy $W_{D,A}$ and battery coherent displacement ergotropy $W_{D,B}$ under strong resonant coupling. Here, $\omega_f=\lambda_{-}$ for (a),(b), $\omega_f=\lambda_{+}$ for (c),(d). $g=0.02\omega$, $F=0.02\omega$, $\gamma_a=0.02\omega$, and other parameters are identical to Fig.\ref{fig1}.}\label{fig2}
\end{figure*}

\begin{figure*}[tb]
\centering
\subfloat
{
\begin{overpic}[width=0.22\textwidth]{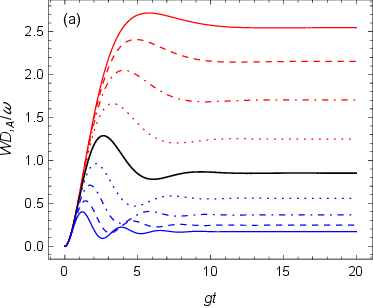}
\end{overpic}
\captionsetup{labelformat=empty}
\label{fig4a}
}
\subfloat
{
\begin{overpic}[width=0.23\textwidth]{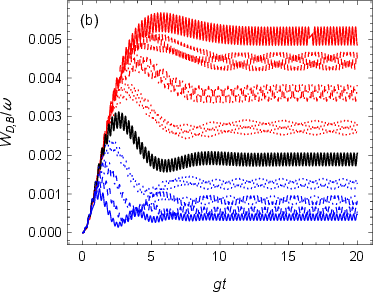}
\end{overpic}
\captionsetup{labelformat=empty}
\label{fig4b}
}
\subfloat
{
\begin{overpic}[width=0.22\textwidth]{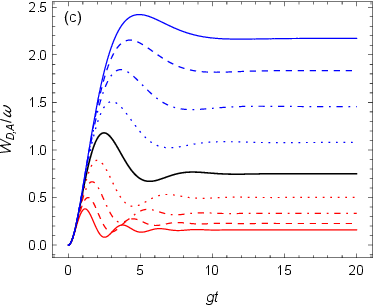}
\end{overpic}
\captionsetup{labelformat=empty}
\label{fig4c}
}
\subfloat
{
\begin{overpic}[width=0.23\textwidth]{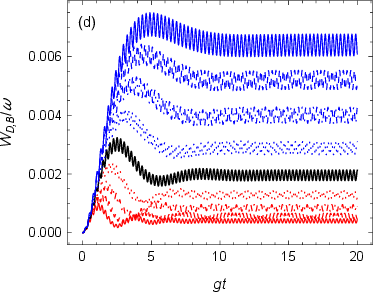}
\end{overpic}
\captionsetup{labelformat=empty}
\label{fig4d}
}
\caption{Evolution of the charger energy $W_{D,A}$ and coherent displacement ergotropy $W_{D,B}$ under strong coupling in the off-resonant bare-frequency regime.
For (a), (b), the driving frequency is fixed at the unperturbed eigenfrequency $\lambda_{-}(\Delta_L=0)$, while for (c), (d) it is fixed at $\lambda_{+}(\Delta_L=0)$.
 Here,  $\omega_a=\frac{2}{3}\omega_b$, and the remaining parameters are the same as in Fig.~\ref{fig2}.}\label{fig3}
\end{figure*}

Compared with Fig.~\ref{fig1}, under a significantly weaker system-bath coupling regime with $g=0.02\omega$, Figs.~\ref{fig2} and~\ref{fig3} present the dynamical evolution of $W_{D,A}$ and $W_{D,B}$ under both resonant and off-resonant driving conditions, respectively.

Under resonant driving (Fig.~\ref{fig2}), the characteristic redshift--blueshift switching behavior remains clearly observable. Nevertheless, in contrast to the results shown in Fig.~\ref{fig1}, both the charger energy $W_{D,A}$ and the coherent displacement ergotropy $W_{D,B}$ exhibit enhanced oscillatory behavior, while the magnitude of the coherent displacement ergotropy $W_{D,B}$ is slightly reduced.

To explicitly address the interplay between stored energy, charging time, and power raised by the Lamb shift, we evaluate the dynamical profile of the average charging power $P_{\mathrm{avg}}(t)$. While a larger battery-like character of the driven supermode enhances the asymptotic ergotropy $W_{D,B}(t \to \infty)$, it simultaneously weakens the direct driving efficiency and extends the characteristic charging time $t_c$. Consequently, the peak charging power $P_{\mathrm{max}}$ does not scale trivially with the maximum long-time ergotropy. As illustrated by the dynamic curves under Protocol I and II, the maximum power $P_{\mathrm{max}}$ occurs during the transient regime ($gt_c \sim \mathcal{O}(1)$) prior to complete dissipative saturation. Tuning $\Delta_L$ to align $\omega_f$ with the renormalized supermodes $\lambda_{\pm}$ allows one to optimize $P_{\mathrm{max}}$ by striking a non-trivial balance between high energy capacity and rapid energy transfer rate. This confirms that the Lamb shift serves not merely as a static storage gating mechanism, but as an active knob for optimizing dynamical charging performance.

This observation clarifies that reducing the inter-subsystem coupling strength $g$ substantially prolongs the characteristic relaxation time scale of the system. Consequently, within the transient timescale of interest ($gt \lesssim 20$), a smaller $g$ lowers the instantaneous rate of coherent energy transfer from the charger to the battery, postponing the accumulation of maximum ergotropy rather than altering the ultimate asymptotic steady-state limit.

Under off-resonant strong driving (Fig.~\ref{fig3}), both the charger output and the coherent displacement ergotropy still maintain a prominent switching effect. However, relative to the previously investigated parameter regimes, the dynamical evolution shows much stronger oscillations before converging to the steady-state regime, and the steady-state oscillation amplitude is also considerably reduced. The enhanced oscillatory dynamics arises from the synergistic interplay between the frequency detuning and the coherent energy exchange mediated by the system-bath coupling interaction.

\subsection{Physical interpretation via normal-mode analysis}

To clarify the dynamical origin of the switching behavior in the charger energy and coherent displacement ergotropy we diagonalize the renormalized Hamiltonian in terms of two hybridized supermodes,
\begin{subequations}\label{15}
\begin{align}
\hat{C}_+ &= \sin\alpha\, \hat{a} + \cos\alpha\, \hat{b}, \label{15a}\\
\hat{C}_- &= \cos\alpha\, \hat{a} - \sin\alpha\, \hat{b}, \label{15b}
\end{align}
\end{subequations}
with
\begin{subequations}\label{16}
\begin{align}
\sin\alpha &=\frac{\omega-\lambda_+}{\sqrt{g^2+(\omega-\lambda_+)^2}},  \label{16a}\\
\cos\alpha &=-\frac{g}{\sqrt{g^2+(\omega-\lambda_+)^2}}. \label{16b}
\end{align}
\end{subequations}
Substituting these operators into Eq.~\eqref{11}, the Hamiltonian becomes

\begin{align}\label{17}
\hat{H}'
=&\lambda_+\hat{C}_+^\dagger \hat{C}_+ + F\sin\alpha \left(e^{i\omega_f t}\hat{C}_+ + e^{-i\omega_f t}\hat{C}_+^\dagger \right) \nonumber\\
 &+\lambda_-\hat{C}_-^\dagger \hat{C}_- + F\cos\alpha \left(e^{i\omega_f t}\hat{C}_- + e^{-i\omega_f t}\hat{C}_-^\dagger \right),
\end{align}
where $\lambda_\pm$ are the eigenfrequencies of the Lamb-shift-renormalized coupling matrix [Eq.~\eqref{13}]. This transformation maps the coherent Hamiltonian onto two non-interacting driven supermodes. In this representation, however, the local reservoir coupling to the charger ($\hat{a}$) inevitably generates non-local cross-dissipation terms between $\hat{C}_+$ and $\hat{C}_-$. Consequently, while the unitary dynamics are decoupled, the supermodes remain dissipatively interconnected through the environment.

This decomposition shows that the energy distribution between the charger and the battery is determined by the modal weights $|\sin\alpha|^2$ and $|\cos\alpha|^2$. However, the supermode formalism itself does not generate the switching effect. The underlying mechanism instead originates from the Lamb-shift-induced renormalization of the eigenfrequencies. Consequently, the observed mode-selective excitation stems from the interplay between the decoupled eigenmode landscape and environment-mediated cross-dissipation. This local-dissipator-induced interference persistently couples the two supermodes during the transient charging process, ultimately shaping the oscillatory profiles of the ergotropy shown in Figs.~\ref{fig1}--\ref{fig3}.

Specifically, the Lamb shift $\Delta_L$ modifies the effective mode frequencies $\lambda_\pm$, thereby changing the resonance condition between the drive $\omega_f$ and the hybridized modes. As $\Delta_L$ varies, the system preferentially absorbs energy through different resonant channels, leading to a redistribution of energy between the charger and the battery [Figs.~\ref{fig1}--\ref{fig3}]. The observed switching in $W_{D,A}$ and $W_{D,B}$ therefore reflects mode-selective excitation within the renormalized energy landscape.

\subsection{Physical distinction between Lamb-shift gating and static detuning}

To further distinguish the present mechanism from trivial detuning in a closed linear system, we emphasize three essential features of the Lamb-shift-induced gating dynamics.

While the effective Hamiltonian in Eq.~(13) suggests that the Lamb shift enters the linearized equations of motion as an effective detuning parameter $\omega_a' = \omega_a + \Delta_L$, the underlying physical mechanism cannot be simplified as a trivial static detuning directly applied to the charger oscillator. It is essential to emphasize three fundamental differences that distinguish environment-induced frequency renormalization from externally imposed frequency shifts:
\begin{enumerate}
    \item \textit{Microscopic constraint via Kramers-Kronig relation:} External static detuning treats frequency shift and subsystem damping as completely independent free parameters. In contrast, the environment-induced Lamb shift $\Delta_L$ and the decay rate $\gamma_a$ are intrinsically bound through the Kramers-Kronig relation [Eqs.~(4a)-(4b)]. Any physical modification of the structured reservoir that shifts $\Delta_L$ fundamentally reshapes the dissipative environment, establishing a self-consistent interplay between coherent hybridization and dissipation that cannot be replicated by manual parametric tuning.
    \item \textit{Non-invasive autonomous gating:} Direct static detuning typically requires invasive external control fields (such as local magnetic flux lines or voltage-induced Stark shifts) coupled directly to the charger, which frequently introduce additional dephasing and thermal noise channels. Conversely, Lamb-shift control operates autonomously and non-invasively through reservoir engineering of the surrounding environment, gating the energy transfer channels without direct perturbation of the subsystem Hamiltonian.
    \item \textit{Non-linear transient ergotropy and cross-dissipation:} Although the field operator expectation values obey linear Heisenberg-Lindblad dynamics [Eqs.~(8a)-(8b)], the extractable work $W_{D,B}(t) = \omega_b |\langle \hat{b}(t) \rangle|^2$ is a non-linear functional of the quantum state. The reservoir coupling generates non-local cross-dissipation between the hybridized supermodes $\hat{C}_\pm$. The Lamb shift modifies the supermode energy splitting $\lambda_\pm$, which directly interferes with these cross-dissipative pathways to reshape the transient peak energy redistribution and charging power before full dissipation sets in.
\end{enumerate}

It is crucial to distinguish the Lamb-shift-induced gating mechanism from ordinary static detuning achieved by directly tuning the bare oscillator frequency $\omega_a$. While both formalisms affect the net detuning in the linearized Heisenberg equations, they possess fundamentally distinct physical origins and thermodynamic implications:

irst, static detuning treats the frequency shift and system dissipation as independent free parameters. In contrast, the Lamb shift $\Delta_L$ and the decay rate $\gamma_a$ are intrinsically tied through the Kramers-Kronig relation [Eqs.~(4a)-(4b)]. Any environmental engineering that modifies $\Delta_L$ inherently reshapes the dissipative landscape $\gamma_a$, leading to a self-consistent, non-trivial interplay between coherent energy exchange and dissipation that cannot be emulated by manual static detuning.

Second, whereas direct frequency tuning requires external, invasive control lines directly coupled to the battery subsystem---often introducing extra dephasing channels---the Lamb-shift control operates non-invasively via environment engineering. By tailoring the spectral density $J(\omega)$ of the surrounding reservoir, energy transfer channels can be gated autonomously.

Third, although the expectation values follow linear equations of motion, the extractable work (ergotropy) $W_{D,B}(t) = \omega_b |\langle \hat{b}(t) \rangle|^2$ is a non-linear functional of the quantum state. The environment-induced shift $\Delta_L$ reconfigures the supermode hybridizations $\lambda_{\pm}$ and cross-dissipation pathways, fundamentally altering the transient energy redistribution peak and charging power before full dissipation sets in.

First, the switching behavior originates from an autonomous environment-induced renormalization rather than externally imposed parameter sweeping. In conventional driven systems, detuning is introduced manually through the drive frequency or cavity parameters. Here, the Lamb shift $\Delta_L$ arises intrinsically from vacuum and thermal fluctuations of the structured reservoir, allowing the QB to self-adjust its resonance structure without real-time external control.

Second, the coherent renormalization and dissipation are microscopically linked. Unlike phenomenological detuning models where frequency mismatch and damping are treated independently, $\Delta_L$ and $\gamma_a$ in our framework originate from the same reservoir spectral function through the Kramers-Kronig relation,
\begin{equation}\label{18}
\Delta_L= \mathcal{P} \int_{0}^{\infty}\frac{J(\omega')}{\omega_0-\omega'}d\omega',
\qquad
\gamma_a = 2\pi J(\omega_0),
\end{equation}
with $J(\omega)$ the environmental spectral density. Consequently, modifying the reservoir simultaneously reshapes both the dissipation landscape and the normal-mode splitting $\lambda_\pm$, which is a genuine open-system effect absent in isolated systems.

Third, the Lamb shift directly influences the thermodynamic performance of the QB. Although the field amplitudes obey linear dynamical equations [Eqs.~\eqref{8a}--\eqref{8b}], the ergotropy $W_{D,B}(t)$ is a nonlinear functional of the battery state [Eq.~\eqref{9}]. The environment-induced switching therefore controls not only the resonance position, but also the conversion efficiency between stored energy and extractable work under dissipation.

These features suggest that the observed switching behavior cannot be fully interpreted as a trivial resonance displacement, but is closely associated with environment-induced mode renormalization and energy redistribution.

\section{Experimental Feasibility}\label{sec:Experimental Feasibility}

To bridge the gap between our theoretical model and practical implementations, we discuss the experimental feasibility of the proposed switchable ergotropy in state-of-the-art quantum platforms\cite{PhysRevApplied.16.044045}. The coupled quantum harmonic oscillator (QHO) model, described by Eq.\eqref{1}, can be naturally realized in circuit quantum electrodynamics (circuit QED) architectures, where superconducting resonators or LC circuits act as the charger and battery subsystems\cite{GROSS1982301,RevModPhys.93.025005}. 
In such architectures, the effective coupling strength $g$ can be wide-rangedly engineered by adjusting the mutual capacitance or inductance between the resonators\cite{Forn-Diaz2019}. The parameters chosen in our simulations (e.g., $g$=0.04$\omega$ and $g$=0.02$\omega$) directly fall into the well-established coupling regime of current circuit QED devices, avoiding the non-RWA pathological anomalies of the deep ultrastrong coupling regime while allowing the Lamb-shift-induced redistribution behavior discussed above.

Furthermore, the environment-induced Lamb shift $\Delta_L$, which serves as the control knob for the switching effect, is a well-established phenomenon in open quantum systems. In circuit QED, the Lamb shift arises from the coupling between a superconducting qubit or resonator and the vacuum fluctuations of a transmission line or a lossy environment. Experimental observations have reported Lamb shifts as large as $1\%$ to $5\%$ of the transition frequency \cite{Fragner2008}, and even larger shifts are achievable in broadband engineered reservoirs. The renormalized resonance condition $\omega'_a = \omega_a + \Delta_L$ can be precisely probed using standard microwave spectroscopy. Moreover, the driving frequency $\omega_f$ of the external laser (or microwave field) is highly tunable in experiments, allowing for the precise targeting of the corrected eigenfrequencies $\lambda_{\pm}$. Given the high quality factors of modern superconducting resonators and the ability to engineer system-environment interference, the proposed frequency correction strategy and the resulting switchable energy distribution may be accessible within current circuit-QED parameter regimes.

\section{Conclusion}\label{sec:Conclusion}
In summary, we have investigated the role of environment-induced frequency renormalization-specifically the Lamb shift-on the transient charging dynamics and energy transfer in an open quantum battery within the Lindblad master equation framework. Our analysis reveals that the Lamb shift redefines the internal resonance condition by renormalizing the coupled eigenfrequencies $\lambda_{\pm}$. This frequency modulation dictates the transient energy partition between the charger and the battery, establishing a trade-off between the achievable coherent displacement ergotropy $W_{D,B}$.

Furthermore, by performing a supermode decomposition, we attribute the observed energy redistribution to mode-selective excitation coupled with bath-mediated cross-dissipation. The resulting shift acts as an intrinsic mechanism that governs whether energy is preferentially stored in the battery or retained in the charger across distinct detuning regimes. These findings demonstrate that coherent environmental corrections cannot be neglected in precision open-system batteries, providing insights for optimizing quantum energy storage in engineered open platforms such as circuit QED.

\section*{Author contributions}
L. Luo conducted all preliminary numerical computations andwrote the original draft of the manuscript. S. C. Zhao designed and conceived the overall research framework of this work, completed datavisualization, analyzed all experimental and numerical results, adjusted relevant numerical calculationdetails, and revised the manuscript. All authors approved the final version of the manuscript.

\section*{Acknowledgment}

This work is supported by the National Natural Science Foundation of China ( Grant Nos. 62065009 and 61565008 ),Yunnan Fundamental Research Projects, China ( Grant No. 2016FB009 ) and the Foundation for Personnel training projects of Yunnan Province, China ( Grant No. KKSY201207068 ).

\section*{Data Availability Statement}

This manuscript has associated data in a data repository.[Authors' comment: All data included in this manuscript are available upon resonable request by contacting with the corresponding author.]
\appendix
\section{Microscopic Derivation of the Lamb Shift}\label{appendix}

In this appendix, we derive the analytical expression of the Lamb shift $\Delta_L$ generated by the structured bosonic reservoir introduced in the main text. Within the Born-Markov approximation, the coherent frequency renormalization originates from the imaginary part of the bath correlation function and is formally expressed as the principal-value integral~\cite{TheoryofOpen1}

\begin{equation}\label{A1}
\Delta_L
=
\mathcal{P}
\int_0^\infty
\frac{J(\omega')}{\omega_a-\omega'}
\,d\omega',
\end{equation}
where $\mathcal{P}$ denotes the Cauchy principal value and the Ohmic spectral density with Lorentz-Drude cutoff is

\begin{equation}\label{A2}
J(\omega')=\frac{\eta}{\pi}\omega'\frac{\omega_c^2}{\omega'^2+\omega_c^2}.
\end{equation}

Substituting Eq.~\eqref{A2} into Eq.~\eqref{A1} yields

\begin{equation}\label{A3}
\Delta_L=\frac{\eta\omega_c^2}{\pi}\mathcal{P}\int_0^\infty\frac{\omega'}{(\omega'^2+\omega_c^2)(\omega_a-\omega')}\,d\omega'.
\end{equation}

To evaluate the integral analytically, we perform the partial-fraction decomposition

\begin{equation}\label{A4}
\frac{\omega'}{(\omega'^2+\omega_c^2)(\omega_a-\omega')}=\frac{A}{\omega_a-\omega'}+\frac{B\omega'+C}{\omega'^2+\omega_c^2},
\end{equation}
with coefficients

\begin{equation}\label{A5}
A=\frac{\omega_a}{\omega_a^2+\omega_c^2},
\qquad
B=\frac{\omega_a}{\omega_a^2+\omega_c^2},
\qquad
C=-\frac{\omega_c^2}{\omega_a^2+\omega_c^2}.
\end{equation}

Equation~\eqref{A3} can therefore be rewritten as

\begin{align}\label{A6}
\Delta_L=\frac{\eta\omega_c^2}{\pi(\omega_a^2+\omega_c^2)}\Bigg[
&
\omega_a\mathcal{P}\int_0^\infty\frac{d\omega'}{\omega_a-\omega'}\nonumber\\
&+
\omega_a\int_0^\infty\frac{\omega' d\omega'}{\omega'^2+\omega_c^2}-\omega_c^2\int_0^\infty\frac{d\omega'}{\omega'^2+\omega_c^2}\Bigg].
\end{align}

The first two terms contain logarithmic divergences individually; however, their Cauchy principal value combination remains strictly convergent. Evaluating the upper and lower limits under a unified high-frequency cutoff prescription yields
\begin{equation}
\mathcal{P}\int_{0}^{\infty}\frac{d\omega'}{\omega_{a}-\omega'} + \int_{0}^{\infty}\frac{\omega' d\omega'}{\omega^{\prime 2}+\omega_{c}^{2}} = \ln\left(\frac{ \omega_{a}}{\omega_{c}}\right).
\end{equation}
The remaining convergent integral evaluates to
\begin{equation}
\int_{0}^{\infty}\frac{d\omega'}{\omega^{\prime 2}+\omega_{c}^{2}} = \frac{\pi}{2\omega_{c}}.
\end{equation}
Collecting all terms, we arrive at the exact microscopic expression for the Lamb shift:
\begin{equation}
\Delta_{L} = \frac{\eta\omega_{c}^{2}}{\pi(\omega_{a}^{2}+\omega_{c}^{2})}\left[\omega_{a}\ln\left(\frac{\omega_{a}}{\omega_{c}} \right) - \frac{\pi}{2}\omega_{c}\right].
\end{equation}
Consequently, the ratio of the coherent Lamb shift to the dissipative decay rate is given by
\begin{equation}
\frac{\Delta_{L}}{\gamma_{a}} = \frac{1}{2\pi}\left[\ln\left( \frac{ \omega_{a}}{\omega_{c}}\right) - \frac{\pi\omega_{c}}{2\omega_{a}}\right].
\end{equation}
This explicit relation underlines the intrinsic interplay between coherent frequency renormalization and environmental dissipation dictated by the microscopic spectral parameters $\eta$ and $\omega_{c}$.

\bibliographystyle{plain}  
\bibliography{references}  
\bibliographystyle{apsrev4-2}

\end{document}